\documentclass[11pt]{article}

\usepackage[reqno]{amsmath}
\usepackage{epsfig}
\usepackage{array}
\usepackage{float}

\def\ltap{\ \raisebox{-.4ex}{\rlap{$\sim$}} \raisebox{.4ex}{$<$}\ }
\def\gtap{\ \raisebox{-.4ex}{\rlap{$\sim$}} \raisebox{.4ex}{$>$}\ }
\newcommand{\deltaatm}{\mbox{$\Delta  m^2_{\mathrm{A}} \ $}}
\newcommand{\deltasol}{\mbox{$ \Delta  m^2_{\odot} \ $}}

\newcommand{\betabeta}{\mbox{$(\beta \beta)_{0 \nu}  $}}

\newcommand{\meff}{\mbox{$\left|  < \! m  \! > \right| \ $}}
\newcommand{\meffmin}{\mbox{$(\left|  
 < \! m  \! > \right|_\mathrm{exp})_{\mbox{}_\mathrm{MIN}} \ $}}
\newcommand{\hbeta}{$\mbox{}^3 {\rm H}$ $\beta$-decay \ }
\newcommand{\eV}{\mbox{$ \  \mathrm{eV} \ $}}
\newcommand{\deltatre}{\mbox{$ \ \Delta m^2_{32} \ $}}
\newcommand{\deltadue}{\mbox{$ \ \Delta m^2_{21} \ $}}

\begin{document}

\hfill{Ref. SISSA 84/2003/EP}

\hfill{Ref. UCLA/03/TEP/26}

\rightline{hep-ph/0310xxx}

\begin{center}
{\bf  Addendum: The SNO Solar Neutrino Data, Neutrinoless Double Beta-Decay 
and Neutrino Mass Spectrum}

\vspace{0.4cm}

S. Pascoli$\mbox{}^a$
~and~
S. T. Petcov$\mbox{}^b$
\footnote{Also at: Institute of Nuclear Research and
Nuclear Energy, Bulgarian Academy of Sciences, 1784 Sofia, Bulgaria}

\vspace{0.2cm}
$\mbox{}^a${\em  Department of Physics \& Astronomy, University of California,
Los Angeles CA 90095-1547, USA}

$\mbox{}^b${\em  Scuola Internazionale Superiore di Studi Avanzati, 
I-34014 Trieste, Italy\\
}
\vspace{0.2cm}   
{\em Istituto Nazionale di Fisica Nucleare, 
Sezione di Trieste, I-34014 Trieste, Italy\\
}
\end{center}

\begin{abstract}

 We update our earlier study in \cite{PPSNO2bb},
which was inspired by the 2002 SNO data,
on the implications of the results of the 
solar neutrino experiments
for the predictions of the effective 
Majorana mass in neutrinoless double beta-decay, \meff.
We obtain predictions for \meff
using the values of the neutrino oscillation parameters,
obtained in the analyzes
of the presently available solar neutrino 
data, including the just published data
from the salt phase of the SNO experiment,
the atmospheric neutrino and CHOOZ data 
and the first data from the KamLAND experiment.
The main conclusion 
reached in \cite{PPSNO2bb}
of the existence
of significant lower bounds on \meff 
in the cases of neutrino mass spectrum 
of inverted hierarchical (IH) and quasi-degenerate (QD)
type is strongly reinforced 
by fact that  combined solar neutrino data
i) exclude the possibility
of  $\cos 2 \theta_\odot = 0$ at more than
5 s.d., ii) determine as a best fit value
$\cos 2 \theta_\odot = 0.40$,
and ii) imply at 95\% C.L. that
$\cos 2 \theta_\odot \gtap 0.22$,
$\theta_\odot$ being the solar neutrino
mixing angle. 
For the IH and QD spectra 
we get using, e.g., the 90\% C.L.
allowed ranges of values
of the oscillation parameters, 
$\meff \gtap 0.010$ eV 
and  $\meff \gtap 0.043$ eV, respectively.
We also comment on the possibility 
to get information on the neutrino mass
spectrum and on the CP-violation in the lepton sector due to
Majorana CP-violating phases.
\end{abstract}

\newpage
\section{Introduction}
\vspace{-0.1cm}

   In the present article we investigate the implications of the 
recently published data from the salt phase of measurements of the
SNO solar neutrino experiment \cite{SNO3} 
for the predictions of the
effective Majorana mass in neutrinoless double beta decay,
\meff (see, e.g., \cite{BiPet87,ElliotVogel02,BPP1}).
We update our earlier similar analysis in \cite{PPSNO2bb},
which was inspired by the 2002 SNO data \cite{SNO2}.
As is well known, the neutrinoless double beta ($\betabeta-$) 
decay, $(A,Z) \rightarrow (A,Z+2) + e^- + e^-$,
is allowed if neutrino mixing,
involving the electron neutrino $\nu_e$,
is present in the weak charged
lepton current and the neutrinos with definite  mass
are Majorana particles (see, e.g., \cite{BiPet87}).
Strong evidences for neutrino mixing, i.e.,
for oscillations of solar electron neutrinos 
$\nu_e$ driven by nonzero neutrino masses 
and neutrino mixing \cite{Pont67}, 
have been obtained in the solar 
neutrino experiments \cite{Cl98,SKsol}:
the pioneering Davis et al. (Homestake) 
experiment \cite{Pont46,Davis68}
and in Kamiokande, SAGE, GALLEX/GNO 
and Super-Kamiokande.
These evidences have been spectacularly reinforced 
during the last two years
by the data from the SNO solar neutrino 
and KamLAND reactor antineutrino experiments.
Under the assumption of CPT-invariance,
the observed disappearance of 
reactor $\bar{\nu}_e$ in the KamLAND experiment
\cite{KamLAND} 
confirmed the interpretation of the solar neutrino data
in terms of $\nu_e \rightarrow \nu_{\mu,\tau}$
oscillations, induced by nonzero 
neutrino masses and 
nontrivial neutrino mixing.
The KamLAND results
practically established the
large mixing angle (LMA)
MSW solution as unique solution
of the solar neutrino problem.

 The combined 2-neutrino oscillation 
analysis of the 
solar neutrino and KamLAND  data
identified two distinct solution sub-regions within 
the LMA solution region - LMA-I,II 
(see, e.g., \cite{fogli,band}).
The best fit values of  the two-neutrino 
oscillation parameters - the solar neutrino mixing
angle $\theta_{\odot}$ 
and the mass squared difference
$\deltasol$, in the two sub-regions - 
LMA-I and 
LMA-II,
read (see, e.g., \cite{fogli}):
$\deltasol^{I} = 7.3 \times 10^{-5}~{\rm eV^2}$,
$\deltasol^{II} = 1.5 \times 10^{-4}~{\rm eV^2}$, 
and $\tan^2 \theta_\odot^{I} = \tan^2 \theta_\odot^{II} = 0.46$.
The LMA-I solution was preferred 
statistically by the data. At 90\% C.L. 
it was found in, e.g., \cite{fogli}: 
\begin{equation}
\deltasol \cong (5.6 - 17) \times 10^{-5}~{\rm eV^2},~~~ 
\tan^2 \theta_\odot \cong (0.32 - 0.72)~.
\label{sol90}
\end{equation}
%
\indent Very recently the SNO collaboration published 
data from the salt phase of the experiment \cite{SNO3}.
For the ratio of the CC and NC event rates, in particular,
the collaboration finds: $R_{CC/NC} = 0.306 \pm 0.026 \pm 0.024$.
Adding the statistical and the systematic errors in quadratures
one gets at 3 s.d.: $R_{CC/NC} \leq 0.41$. 
As was shown in \cite{MarisSP02},
an upper limit of $R_{CC/NC} < 0.5$ implies a significant
upper limit on $\deltasol$ smaller than $2\times 10^{-4}~{\rm eV^2}$.
In the case of interest, $R_{CC/NC} \leq 0.41$,
one finds using the results from \cite{MarisSP02}:
$\deltasol \ltap 1.5\times 10^{-4}~{\rm eV^2}$.
Thus, the new SNO data on 
$R_{CC/NC}$ implies stringent constraints 
on the LMA-II solution. A combined statistical analysis 
of the data from the solar neutrino and KamLAND experiments,
including the latest SNO results,
shows \cite{SNO3BCGPR} that the LMA-II solution is allowed 
only at 99.13\% C.L. 
Furthermore, the data have substantially 
reduced the maximal allowed value 
of $\sin^2 \theta_\odot$,
excluding the possibility of maximal mixing 
\footnote{
The future data from SNO on the day-night effect
and the spectrum of $e^-$ from the CC reaction, and 
the future high statistics data from KamLAND,
in principle, can resolve completely 
the LMA-I - LMA-II solution ambiguity 
and can constrain further
the solar neutrino mixing angle (see 
\cite{MarisSP02,MarisSP00DN,SNO3BCGPR} 
and the references quoted therein).}
at 5.4 s.d. This has
very important implications for the predictions of \meff
\cite{PPSNO2bb,PPW,BGKP96}.

   A 3-$\nu$ oscillation 
analyzes of all available data, 
including the latest SNO results,
were performed in \cite{SNO3BCGPR,Valle2003}
~\footnote{Combined 2-neutrino oscillation
analyzes of the solar neutrino and KamLAND data
were completed earlier in refs.~\cite{bala2003},
and in \cite{SNO3Milano}.}.
It was found, in particular,
in \cite{SNO3BCGPR} that at 90\%~C.L.,
\begin{eqnarray}
0.23 \ltap \sin^2 \theta_\odot \ltap 0.38 \ \ \ \  
\text{for}\ \ \ \ \sin^2 \theta = 0.0 ,\\
0.25 \ltap \sin^2 \theta_\odot \ltap 0.36 \ \ \ \  
\text{for}\ \ \ \ \sin^2 \theta = 0.04, 
\end{eqnarray}
%
\noindent where $\theta$ is the angle limited by the
CHOOZ and Palo Verde experiments \cite{CHOOZPV}.
The best fit (BF) values in both cases read
$\sin^2 \theta_\odot = 0.30$.
The allowed values of \deltasol
in the LMA-I region have not changed considerably
and are given at 90\%~C.L. by \cite{SNO3BCGPR}:
\begin{eqnarray}
5.6 \times 10^{-5} \ \eV^2  \ltap \deltasol 
\ltap 9.2 \times 10^{-5} \ \eV^2  \ \ \ \  
\text{for} \ \ \ \  \sin^2 \theta = 0.0, \\
6.1 \times 10^{-5} \ \eV^2  \ltap \deltasol \ltap 
8.5 \times 10^{-5} \ \eV^2  \ \ \ \  \text{for}  \ \ \ \ \sin^2 \theta = 0.04. 
\end{eqnarray}
%
\noindent The $\deltasol$ best fit value is practically
the same for the two values of $\sin^2 \theta$:
$\deltasol \cong 7.2  \times 10^{-5} \ \eV^2$.

  There are also strong evidences for
oscillations of the atmospheric 
$\nu_{\mu}$ ($\bar{\nu}_{\mu}$) from
the observed Zenith angle dependence of the multi-GeV 
$\mu$-like events in the SuperKamiokande experiment
\cite{SKatmo03}.
The combined analysis~\cite{FogliatmKamL} 
of atmospheric neutrino data and 
the data from the 
K2K long base line accelerator experiment \cite{K2K},
shows that at 90\%~C.L. 
the neutrino mass squared difference 
responsible for the atmospheric
neutrino oscillations lies in the interval
\begin{equation}
2.0 \times 10^{-3} \ \eV^2 \ltap \deltaatm \ltap 3.2 \times 10^{-3} \ \eV^2.
\label{atmoBari}
\end{equation}
%
\noindent The $\deltaatm$ best fit value 
found in \cite{FogliatmKamL} reads:
 $\deltaatm|_\mathrm{BF} = 2.6 \times 10^{-3} \ \eV$.
Let us note that the preliminary results 
of an improved analysis of the
SK atmospheric neutrino data, performed recently by the
SK collaboration, gave \cite{SKatmo03}
\begin{equation} 
1.3 \,\times\, 10^{-3}\,\mbox{eV}^2\,\ltap\,
|\deltaatm|\, \ltap\,3.1\, \times\,10^{-3}\,\mbox{eV}^2~,
 ~~~~~90\%~{\rm C.L.},
\label{atmo03a}
\end{equation}
%
\noindent with best fit value $|\deltaatm| = 2.0\times 10^{-3}$ eV$^2$.
Adding the K2K data~\cite{K2K}, 
the authors \cite{Fogliatm0308055}
find the same best fit value and
\begin{equation} 
1.55 \,\times\, 10^{-3}\,\mbox{eV}^2\,\ltap\,
|\deltaatm|\, \ltap\,2.60\, \times\,10^{-3}\,\mbox{eV}^2~,
 ~~~~~90\%~{\rm C.L.}
\label{atmo03}
\end{equation}

  The last parameter relevant
for our further discussion is 
the neutrino mixing angle $\theta$,
limited by CHOOZ and Palo Verde
experiments. The precise limit on 
$\theta$ is $\deltaatm-$ dependent (see, e.g, 
\cite{BNPChooz}).
For the values of \deltaatm found in \cite{FogliatmKamL} 
(see eq. (\ref{atmoBari})), 
the upper bound  on $\sin^2\theta$
at 90\% (99.73\%) C.L. reads:
$\sin^2 \theta < 0.03~(0.05)$.
Using the latest SK preliminary result, 
one gets at 90\% (99.73\%) C.L.
from  a combined 3-$\nu$
oscillation analysis of the solar neutrino,
CHOOZ and KamLAND data \cite{SNO3BCGPR}:
\begin{equation}
\sin^2 \theta < 0.047~(0.074).
\end{equation}
%
\noindent

   Under the assumptions of 3-neutrino mixing,
for which we have compelling evidences from the experiments with
solar and atmospheric neutrinos and from the KamLAND 
experiment,
of massive Majorana neutrinos and
of  \betabeta-decay generated
{\it only by the (V-A) charged current 
weak interaction via the exchange of the three
Majorana neutrinos  $\nu_j$},
the effective Majorana mass 
in \betabeta-decay
of interest is given by
(see, e.g., \cite{BiPet87,BPP1}):
\begin{equation}
\meff  = \left| m_1 |U_{\mathrm{e} 1}|^2 
+ m_2 |U_{\mathrm{e} 2}|^2~e^{i\alpha_{21}}
 + m_3 |U_{\mathrm{e} 3}|^2~e^{i\alpha_{31}} \right|~,
\label{effmass2}
\end{equation}
\noindent where 
$U_{ej}$, $j=1,2,3$, are the elements of the 
first row of 
the Pontecorvo-Maki-Nakagawa-Sakata (PMNS)
neutrino mixing matrix \cite{BPont57} $U$,
$m_j > 0$ is the mass of the Majorana neutrino $\nu_j$,
and $\alpha_{21}$ and $\alpha_{31}$ 
are two Majorana CP-violating phases
\cite{BHP80,Doi81}.
One can express 
\cite{SPAS94} the masses $m_{2,3}$ 
and the elements $U_{ej}$
respectively in terms of $m_1$, $\deltasol$
$\deltaatm$, and of
$\theta_\odot$, $\theta$ (see further).

   In ref.~\cite{PPSNO2bb} we have analyzed 
the implications of the  
results of the 
solar neutrino experiments, including the 2002
SNO data, which favored the LMA MSW solution of the 
solar neutrino problem with
$\tan^2\theta_{\odot} < 1$, 
for the predictions of \meff.
Neutrino mass spectra with normal mass hierarchy, 
with inverted hierarchy and of quasi-degenerate type 
were considered (see, e.g., \cite{BPP1}).
From the fact that $\cos 2 \theta_\odot \geq 0.26$,
which followed (at 99.73\% C.L.)
from the analysis of the solar neutrino data 
performed in \cite{SNO2},
we found significant lower bounds on
\meff in  the cases of neutrino mass spectrum 
of quasi-degenerate and 
inverted hierarchical type:
$\meff \gtap 0.03$ eV and                          
$\meff \gtap 8.5\times 10^{-3}$ eV, respectively.
We have also found that if the 
neutrino mass spectrum were hierarchical (with inverted
hierarchy), \meff were limited from above: 
$\meff  \ltap 8.2 \times 10^{-3}~(8.0 \times 10^{-2})$ eV.
These results led us to conclude
that a measured value of 
$\meff \gtap 10^{-2}$ eV
could provide fundamental
information on the type of
the neutrino mass spectrum.
It was also shown that 
such a result could provide  
a significant upper limit 
on the mass of the lightest
neutrino $m_1$ as well.
In refs.~\cite{PPRSNO2bb,PPR2} 
the possibilities to 
determine the type of neutrino mass spectrum 
and to get information of 
CP-violation in the lepton sector,
associated with Majorana neutrinos 
from a measurement of \meff
have been studied in greater detail.
In these articles, in particular,
the uncertainty in the determination
of \meff due to an imprecise knowledge
of  the relevant nuclear matrix elements
and prospective experimental errors 
in the measured values of the 
neutrino oscillation parameters,
entering into the expression of \meff,
were taken into account.
In \cite{PPRSNO2bb} the results of analyzes 
of the solar and atmospheric
neutrino and the first KamLAND data were used
in obtaining predictions for \meff
\footnote{A concise discussion of the relevant formalism,
of the status of the searches for \betabeta-decay and
of the physics potential of the \betabeta-decay
experiments as well as of the predictions 
for \meff before the publication of the latest
SNO data, can be found in \cite{PPVenice03}.}. 

  In the present Addendum, we update the predictions for
\meff derived in ref.~\cite{PPSNO2bb} and the conclusion
reached in the indicated article  
by taking into account the implications of the
recently announced data from the salt phase 
measurements of the SNO experiment.
We also comment very briefly how the results
obtained in \cite{PPRSNO2bb,PPR2} 
are modified.

\section{Predictions for the Effective Majorana Mass Parameter \meff}


   The predicted value 
of \meff{} depends in the 
case of $3-\nu$ mixing on 
the oscillation parameters \deltaatm$\!\!$,
$\theta_{\odot}$, $\Delta m^2_{\odot}$ and 
$\theta$ (see, e.g., \cite{BPP1}). 
Following \cite{PPSNO2bb}, we will use 
the convention in which  $m_1 < m_2 < m_3$. 
In this convention one has
$\deltaatm \equiv \Delta m^2_{31}$.
In what regards $\deltasol$, there 
are two possibilities:
$\deltasol \equiv \Delta m^2_{21}$, 
or $\deltasol \equiv \Delta m^2_{32}$.
The former corresponds to neutrino
mass spectrum with normal hierarchy,
while the latter - to neutrino mass spectrum 
with inverted hierarchy.
Obviously, $m_3 = \sqrt{m_1^2 + \deltaatm}$.
If $\deltasol \equiv \Delta m^2_{21}$, one has   
$m_2 = \sqrt{m_1^2 + \deltasol}$
and the following relations are valid:
$|U_{\mathrm{e} 1}|^2 = \cos^2\theta_{\odot} (1 - |U_{\mathrm{e} 3}|^2)$, 
$|U_{\mathrm{e} 2}|^2 = \sin^2\theta_{\odot} (1 - |U_{\mathrm{e} 3}|^2)$,
and  $|U_{\mathrm{e} 3}|^2 \equiv \sin^2\theta$.
In the case of neutrino mass spectrum with inverted hierarchy,
we have $m_2 = \sqrt{m_1^2 + \deltaatm - \deltasol}$ and
$|U_{\mathrm{e} 2}|^2 = \cos^2\theta_{\odot} (1 - |U_{\mathrm{e} 1}|^2)$, 
$|U_{\mathrm{e} 3}|^2 = \sin^2\theta_{\odot} (1 - |U_{\mathrm{e} 1}|^2)$,
$|U_{\mathrm{e} 1}|^2 \equiv \sin^2\theta$.
Given \deltasol$\!\!$, \deltaatm$\!\!$, 
$\theta_{\odot}$ and
$\sin^2\theta$, the value of \meff{} 
depends strongly on the type of the
neutrino mass spectrum and 
on the value of 
the lightest neutrino mass, $m_1$,
as well as on the values of the two
Majorana CP-violation phases
present in the PMNS matrix,
$\alpha_{21}$ and $\alpha_{31}$ 
(see eq.\ (\ref{effmass2})).

   As in ref.~\cite{PPSNO2bb}, 
we consider the predictions of \meff
for the following three specific types of 
neutrino mass spectrum:
i) normal {\it hierarchical} (NH), corresponding 
to a spectrum with normal hierarchy and $m_1 \ll m_{2,3}$,
ii) inverted {\it hierarchical} (IH), 
characterized by inverted hierarchy and  
$m_1 \ll m_{2,3}$,
and iii) quasi-degenerate spectrum (QD)
which is realized if $m_1 \cong m_2 \cong m_3 \equiv m_0$ and
$m_{1,2,3}^2 \gg \deltaatm\!\!, \deltasol$.
Let us note that  in the case 
of the QD spectrum,
~\meff{} is essentially independent of
\deltaatm and \deltasol, and,
as long as the Majorana CP-violation phases
$\alpha_{21}$ and $\alpha_{31}$
are not constrained,
the two possibilities, 
$\deltasol \equiv \deltadue$
and $\deltasol \equiv \deltatre \!\!$, 
lead {\it effectively} 
to the same predictions for 
the allowed range of values of \meff{}. 

   In Tables 1 and 2  we give the 
i) maximal predicted value of \meff{} 
in the case of NH neutrino mass spectrum,
ii) the minimal and maximal  values of \meff{} 
for the IH spectrum,
and iii) the minimal value of \meff{} 
for the  QD spectrum. The indicated values of 
\meff are calculated 
for the best fit values and for the 90\%~C.L.\ 
allowed ranges of \deltaatm from 
\cite{FogliatmKamL} and
\cite{Fogliatm0308055},
and of $\sin^2\theta_{\odot}$ and \deltasol~
in the LMA-I solution region, 
obtained in the analysis 
in ref.~\cite{SNO3BCGPR}.
\begin{table}[ht]
\begin{center}
\begin{tabular}{|c|c|c|c|c|} 
\hline
\rule{0pt}{0.5cm} $\sin^2 \theta$
& $\meff_{\rm max}^{\rm NH}$ & $\meff_{\rm min}^{\rm IH}$ &  $\meff_{\rm max}^{\rm IH}$ &
$\meff_{\rm min}^{\rm QD} $  
\\ \hline \hline
0.0   & 2.6 (2.6)   &  19.9 (17.3)    &  50.5 (44.2)   & 79.9 \\ \hline
 0.02 & 3.6 (3.5)   &  19.5 (17.0)    &  49.5 (43.3)   & 74.2 \\ \hline
 0.04 & 4.6 (4.3)   &  19.1 (16.6)    &  48.5 (42.4)   & 68.5 \\ \hline
\end{tabular}
\caption{\label{tab:bf} 
The maximal values of  
\meff{} (in units of $10^{-3}$ eV)
for the NH  and IH spectra, and the minimal values of 
\meff{} (in units of $10^{-3}$ eV) for the IH and QD spectra,
for the best fit values of the oscillation parameters and 
$\sin^2\theta = 0.0$, $0.02$ and 
$0.04$.
The results for the NH and IH spectra 
are obtained for 
$\deltaatm|_\mathrm{BF}= 2.6 \times 10^{-3} \ \eV^2~
(2.0 \times 10^{-3} \ \eV^2 -~{\rm values~in~brackets})$ and
$m_1 = 10^{-4}$ eV, while
those for the QD spectrum correspond to 
$m_0 = 0.2$ eV\@.
}
\end{center}
\end{table}
\begin{table}[ht]
\begin{center}
\begin{tabular}{|c|c|c|c|c|} 
\hline
\rule{0pt}{0.5cm} $\sin^2 \theta$
& $\meff_{\rm max}^{\rm NH}$ & $\meff_{\rm min}^{\rm IH}$ &  $\meff_{\rm max}^{\rm IH}$ &
$\meff_{\rm min}^{\rm QD} $  
\\ \hline \hline
0.0  & 3.7 (3.7) & 10.1 (8.7) & 56.3 (50.6) & 47.9 \\ \hline
0.02 & 4.7 (4.6) &  9.9 (8.6) & 55.1 (49.6) & 42.8 \\ \hline
0.04 & 5.5 (5.3) & 11.4 (9.9) & 54.0 (48.6) & 45.4 \\ \hline
\end{tabular}
\caption{\label{tab:90} 
The same as in Table 1 but 
for the 90\% C.L. allowed values of 
$\deltasol$ and $\theta_\odot$ obtained in \cite{SNO3BCGPR}, 
and of \deltaatm given in
eq. (6) (eq. (8) - results in brackets).
}
\end{center}
\end{table}
%
In Figs. 1 and 2 we show 
the allowed ranges of values of \meff as a function
of $m_1$ for the cases of spectrum with normal and 
inverted hierarchy. The predictions for \meff
are obtained 
by using the best fit (Fig. 1), and the allowed
at 90\%~C.L. (Fig. 2), values of
$\deltasol$, $\theta_\odot$ and \deltaatm
from refs. ~\cite{SNO3BCGPR} and 
\cite{Fogliatm0308055}
and for three fixed values of $\sin^2 \theta$.

  Let us recall that   
$\meff_{\rm min}^{\rm IH,QD}$ 
are given approximately 
by \cite{PPSNO2bb} (see also \cite{PPW,BGKP96}):
\begin{equation}
\meff_{\rm min}^{\rm IH} \cong
\sqrt{\deltaatm} \cos^2\theta~|\cos2\theta_{\odot}|,
\end{equation}
%
\begin{equation}
\meff_{\rm min}^{\rm QD} \cong 
m_0\left |\cos^2\theta~\cos2\theta_{\odot} - \sin^2\theta \right |.
\end{equation}
%
\noindent According to \cite{Fogliatm0308055,SNO3BCGPR},
we have at 99.73\% C.L.:
$\deltaatm \gtap 1.1\times 10^{-3}~{\rm eV^2}$ and
$\sin^2\theta < 0.074$. The combined analysis
of the solar neutrino data, including the latest
SNO results, shows \cite{SNO3,SNO3BCGPR} 
that i) the possibility of  $\cos2\theta_{\odot} = 0$
is excluded at more than 5 s.d.,
ii) the best fit value of 
$\cos2\theta_{\odot}$ is 
$\cos2\theta_{\odot} = 0.40$, and
iii) that at 95\% C.L. one has
for $\sin^2\theta = 0~(0.04)$,
$\cos2\theta_{\odot} \gtap 0.22~(0.24)$.
These new results firmly establish
the existence of significant
lower bounds on \meff 
in the cases of IH and QD neutrino mass spectra.
That has fundamental implications
for the searches for \betabeta-decay.

  A comparison of the results for \meff, obtained 
using the best fit values of 
$\deltasol$, $\theta_\odot$ and \deltaatm,
with those reported in 
refs.~\cite{PPSNO2bb,PPR2}
shows that the predictions for \meff
did not change considerably.
This is a consequence of the fact that 
the best fit values of \deltasol and 
$\sin^2 \theta_\odot$
are not very different from the values
used as input in refs.~\cite{PPSNO2bb,PPR2}.
With the recent preliminary result 
$\deltaatm|_\mathrm{BF} = 2.0 \times 10^{-3} \ \eV^2$
of the improved analysis of 
the SK atmospheric neutrino data \cite{SKatmo03} 
one gets somewhat smaller values of \meff
in the case of IH spectrum.
For $\sin^2 \theta \sim 0$, for instance, one finds
$17 \times 10^{-3} \ \eV \ltap \meff^\mathrm{IH}
\ltap 44  \times 10^{-3} \ \eV$. The ranges of \meff
in the NH and QD spectra depend weakly 
on \deltaatm in this case
\footnote{The effective Majorana mass
\meff for the NH spectrum practically does not depend on \deltaatm
if $\sin^2 \theta$ is sufficiently small, so that
$\sqrt{\deltasol} \sin^2 \theta_\odot \gg
\sqrt{\deltaatm} \sin^2 \theta$.
Even for $\sin^2 \theta$ close to its
90\% C.L. upper limit of 0.047,
the change in $\meff_{\rm max}^{\rm NH}$
due to the lower value of $\deltaatm|_\mathrm{BF}$
amounts at most to 6~\%.}
and therefore 
$\meff_{\rm max}^{\rm NH}$ and $\meff_{\rm min}^{\rm QD}$ 
reported in Table 1 are essentially the same as 
the ones reported in refs.~\cite{PPSNO2bb,PPR2}.

 According to the  3-$\nu$ combined analysis 
of the solar neutrino and KamLAND and CHOOZ data 
\cite{SNO3BCGPR},
for $\sin^2\theta = 0~(0.04)$,
the lower bound on $\cos 2 \theta_\odot$
at 90\%~C.L. reads:
$\cos 2 \theta_\odot|_{\mbox{}_\mathrm{MIN}} 
\gtap 0.24~(0.28)$.
Therefore the main conclusion 
that in the case of the IH and the QD spectra
there exist 
significant lower bounds on \meff ~\cite{PPSNO2bb}, 
not only still holds, but is 
considerably strengthened, 
as we have already emphasized.
More specifically, 
in the case of IH spectrum
we get for $\sin^2\theta = 0~(0.04)$
$\meff_{\rm min}^{\rm IH} = 0.010~(0.011) \ \eV$ if we use
eq. (6), and 
$\meff_{\rm min}^{\rm IH} = 0.0087~(0.0099)\ \eV$
utilizing the preliminary result given in eq. (8).

   In the case of QD neutrino mass spectrum,
a larger lower bound on $\cos 2 \theta_\odot$
implies a larger value of $\meff_{\rm min}^{\rm QD}$,
which now reads:
$\meff_{\rm min}^{\rm QD} \sim 0.043-0.048 \ \eV$.
This should be compared with the value found in \cite{PPR2}:
$\meff_{\rm min}^{\rm QD} \sim 0.03 \ \eV$.

 Let us note that for the 
presently allowed at 90\%~C.L. 
values of \deltasol,
$\theta_\odot$ and \deltaatm,
and for $\sin^2 \theta \gtap 0.03$,
there exists a  lower bound on \meff 
in the case of NH spectrum, provided 
$m_1 << 10^{-3}$ eV: one finds 
$\meff^\mathrm{NH}_{\mbox{}_\mathrm{max}} 
\gtap \mathrm{few} \times 10^{-4} \ \eV$.
A complete cancellation 
of the different terms contributing to \meff
is allowed in this case only if
\begin{equation}
\sin^2 \theta > \sqrt{\frac{(\deltasol)_{\mbox{}_\mathrm{MIN}}}
{(\deltaatm)_{\mbox{}_\mathrm{MAX}}}} 
(\sin^2 \theta_\odot)_{\mbox{}_\mathrm{MIN}}.
\end{equation}
%
Given the 90\% C.L. allowed ranges of the oscillation
parameters entering into the above inequality, 
the ``cancellation'' condition can be
satisfied for $\sin^2 \theta > 0.038$. 
Lower allowed values of \deltaatm
or a more stringent upper 
limit on $\sin^2 \theta$
would further strengthen this result. 
It should be emphasized, however,
that one can have $\meff \cong 0$
even for $\sin^2 \theta \ltap 0.030$
if $m_1$ is sufficiently large  
(see Figs. 1 and 2).
More generally,
since at present the two 
Majorana CP-violation
phases $\alpha_{21}$ and $\alpha_{31}$
are not constrained and 
the existing upper bounds 
on $m_1$ are not sufficiently stringent,
one can always have $\meff = 0$ 
in the case of neutrino
mass spectrum with normal hierarchy
\cite{PPW}.

  The 95\% C.L. allowed ranges of 
\deltasol and  $\theta_\odot$ differ only 
marginally from those derived at
90\% C.L., while the upper limit on
$\sin^2\theta$ changes from 0.047 to
0.053 \cite{SNO3BCGPR}.
The intervals of allowed values 
of \deltaatm at 95\% C.L.,
according to the older and the latest
analyzes \cite{FogliatmKamL} and 
\cite{Fogliatm0308055} 
do not differ considerably 
from the 90\% C.L. ones, eqs. (6) and (8), and
are given respectively by
$\deltaatm = (1.8 - 3.4)\times 10^{-3}~{\rm eV^2}$ and 
$\deltaatm = (1.4 - 2.8)\times 10^{-3}~{\rm eV^2}$.
These results imply that the predictions
for \meff, obtained using the 95\% C.L.
allowed ranges of the neutrino oscillation 
parameters, differ insignificantly
from the predictions based on
the 90\% C.L. ranges of the parameters,
which are illustrated in Table 2 and
Fig. 2.  

    Similar conclusion about the existence of
a significant and robust lower bound on \meff , 
$\meff \gtap 10^{-2}$ eV, in the case of 
IH neutrino mass spectrum,
has been reached also in \cite{Carlosbb03}, where 
a $\chi^2-$method of analysis was employed.

\section{Constraining the Type of Neutrino Mass Spectrum 
and/or CP-violation Associated with Majorana Neutrinos}


    In refs.~\cite{PPRSNO2bb,PPR2}
the possibilities to discriminate
between the different types of 
neutrino mass spectrum 
and to get information 
about CP-violation associated with Majorana neutrinos
by means of a measurement of
\meff  were studied.
The uncertainty 
in the relevant nuclear matrix elements 
and prospective experimental errors 
in  the values of the oscillation parameters,
in \meff, and for the case of QD spectrum - in $m_1$,
were taken into account.
Here we update the results of \cite{PPRSNO2bb,PPR2}
using the same notations.
We denote by \meffmin the value of \meff
obtained from a measurement of the \betabeta-decay
half life-time of a given nucleus by
using the largest physical nuclear matrix element,
and by $\zeta$  the ``theoretical
uncertainty'' in \meff due to the imprecise
knowledge of the nuclear matrix element.
Thus,  an experiment measuring 
a \betabeta-decay half-life time will determine
a range of values of \meff corresponding to
\begin{equation}
\meffmin - \Delta \leq \meff \leq \zeta
( \meffmin + \Delta),
\end{equation}
%
where $\Delta$ denotes the experimental error 
in the measurement of \meff.
A part of the analysis in \cite{PPRSNO2bb,PPR2} 
was performed by using
the best fit values of the solar and the atmospheric
neutrino oscillation parameters
and including 1 s.d. (3 s.d.) uncertainties
of 5\% (15\%) on $\tan^2 \theta_\odot$ 
and \deltasol,
and of 10\% (30\%) on \deltaatm.
We follow  \cite{PPRSNO2bb,PPR2} 
in the choice for the ranges of the 
indicated parameters.
In Table 3 we report the predicted
i) maximal  value of \meff{} 
in the case of NH spectrum,
ii) the minimal and maximal  values of \meff{} 
for the IH neutrino mass spectrum,
and iii) the minimal value of \meff{} 
for the  QD spectrum.

\begin{table}[ht]
\begin{center}
\begin{tabular}{|c|c|c|c|c|} 
\hline
\rule{0pt}{0.5cm} $\!\! \!\sin^2 \theta \!\!$
& $\meff_{\rm max}^{\rm NH}$ & $\meff_{\rm min}^{\rm IH}$ &  $\meff_{\rm max}^{\rm IH}$ &
$\meff_{\rm min}^{\rm QD} $  
\\ \hline \hline
\!\! 0.0 \!\! & \!\!\!   2.8 (3.1)  [2.8 (3.1)] \!\!\!  
& \!\!\!  17.8 (13.8) [15.5 (12.0)] \!\!  &\!\!\!  53.0 (57.8) [46.4 (50.6)]\!\!\!  
& \!\!\!  78.0 (69.4) [77.6 (69.1)]\!\! \\ \hline
\!\! 0.02 \!\! & \!\!\!    3.8 (4.2) [3.7 (4.1)]\!\!\!  
& \!\!\!  17.4 (13.5) [15.2 (11.7)]\!\!  &\!\!\!  52.0 (56.6) [45.5  (49.6)]\!\!\!  
& \!\!\!  72.5 (64.1) [72.0 (63.7)]\!\! \\ \hline
\!\! 0.04 \!\! & \!\!\!   4.8 (5.3) [4.6  (5.0)] \!\!\!  
& \!\!\!  17.0 (13.3) [14.8 (11.5)]\!\!  &\!\!\!   50.9 (55.5) [44.5  (48.6)]\!\!\!  
& \!\!\!  66.9 (58.7) [66.4 (58.4)]\!\! \\ \hline
\end{tabular}
\caption{\label{tab:1sd} 
The maximal values of  \meff{} 
for the NH  and the IH spectrum and the minimal values of 
\meff{} for the IH and QD spectra (in units of $10^{-3}$ eV),
obtained by using the  best fit values 
of the solar and the atmospheric
neutrino oscillation parameters
and including 1 s.d. (3 s.d.) uncertainties
of 5\% (15\%) on $\tan^2 \theta_\odot$ 
and \deltasol, and of 10\% (30\%) on \deltaatm.
Results for $\sin^2\theta = 0.0$, $0.02$ and 
$0.04$ are shown.
Two values of $\deltaatm|_{\mathrm{BF}}$ are used:
$\deltaatm|_{\mathrm{BF}} = 2.6 \times 10^{-3} \ \eV^2  
~[2.0 \times 10^{-3} \ \eV^2]$.
The results for the NH and IH spectra 
are obtained for $m_1 = 10^{-4}$ eV, while
those for the QD spectrum correspond to 
$m_0 = 0.2$ eV\@.
}
\end{center}
\end{table}

 In order to be possible to 
discriminate between the NH and 
the IH spectrum, the following inequality 
should be fulfilled \cite{PPRSNO2bb}: 
$ \zeta \meff_{\rm max}^{\rm NH} <  \meff_{\rm min}^{\rm IH}$.
The latter implies an upper limit on $\zeta$.
For the currently favored 
values of the neutrino oscillation 
parameters and $\sin^2\theta \ltap 0.03$, 
the NH spectrum can be distinguished from the IH one  
even if $\zeta \sim 3$. 
If $\sin^2\theta \gtap 0.03$, $\meff_{\rm max}^{\rm NH}$
can be larger, as Table 3 illustrates,
and somewhat smaller values of $\zeta$ 
could be required.
Similarly, since $\meff_{\rm min}^{\rm IH} \sim
\sqrt{\deltaatm} | \cos 2 \theta_\odot |$, a shift of
\deltaatm 
to smaller values
would lead to stronger constraints on $\zeta$.
In the ``worst possible case'' in which
we allow a 3 s.d. error on \deltasol, \deltaatm 
and $\tan^2 \theta_\odot$,
and use the best fit values 
of the solar neutrino oscillation parameters, 
$\deltaatm = 2.0 \times 10^{-3} \ \eV^2$ and 
$\sin^2 \theta=0.04$,
it is necessary to have $\zeta < 2.3$.

The possibility to discriminate 
between the NH and the QD spectra   
leads to a less stringent condition
on $\zeta$ than the condition
permitting to
distinguishing between the NH and the IH spectra:
the former is satisfied 
even for values of $\zeta$ exceeding 3. 
The IH spectrum can be distinguished 
from the QD spectrum 
only if $\zeta \ltap 1.5$~
\footnote{
The upper bound on $\zeta$ varies
from 1.1 to 1.6 according to the different values
of \deltaatm and $\sin^2 \theta$ (see Table 3).
The constraint is less stringent for smaller values of
 \deltaatm and $\sin^2 \theta$.},
unless additional information on neutrino masses 
is provided by the \hbeta experiments 
and/or cosmological and astrophysical measurements.

   We update also 
the conditions on $\zeta$ and $\Delta$
which would permit to rule out, or establish,
the NH, IH and the QD mass spectrum. 
The next generation of 
\betabeta-decay experiments are planned
to reach a sensitivity of 
$\meff \sim 0.01 - 0.03\ \eV$.
The QD mass spectrum can be ruled out
if $\zeta < \meff_{\rm min}^{\rm QD} / ( \meffmin + \Delta)$.
For the prospected sensitivity of $\meff \sim 0.01 \ \eV$
and  $\Delta \sim 0.01 \ \eV$,
the requirement on $\zeta$ reads $\zeta < 3$.
In the less favorable case in which 
$(\meffmin + \Delta) \sim 0.05 \ \eV$,
an extremely accurate knowledge of the nuclear matrix elements,
i.e. $\zeta \sim 1$, would be necessary.
In order to establish that the spectrum is 
of the QD type,
the following condition has to be fulfilled:
$(\meffmin -\Delta) \geq 0.2 \ \eV$,
as it implies that $m_0 \geq 0.2 \ \eV$.
This requirement is satisfied if, e.g., 
the measured  
$(\meff_{\rm exp})_{\mbox{}_{\rm MIN}}\sim 0.3$ eV
with an experimental error 
$\Delta \ltap 0.1$ eV\@. 
The NH spectrum can be excluded 
provided the measured value of \meff{} 
is larger than the predicted upper limit
on \meff for this type of spectrum,
$\meffmin - \Delta >\meff_{\rm max}^{\rm NH}$.
Given the expected sensitivities
on \meff,
this condition can be realized if,
e.g., $\meffmin$ is at least 3-11 times larger 
than $\meff_{\rm max}^{\rm NH}$,
i.e. ($\meffmin \sim 0.015 - 0.035$),
and the experimental error amounts to at most 
$\Delta \sim 0.01 - 0.03 \ \eV$.
A larger measured value of \meffmin
would allow to exclude the NH spectrum
even for larger values of $\Delta$.
The IH spectrum can be ruled out 
if $\meffmin - \Delta > \meff_{\rm max}^{\rm IH}$.
For an experimental error on \meff,
$\Delta \simeq 0.01;~ 0.03;~0.05 \ \eV$,
this condition is satisfied 
if $\meffmin > 0.06;~ 0.08;~ 0.1 \ \eV$.
Establishing that the 
spectrum is of the IH type
is quite demanding and
requires a measurement 
of \meff{} with an error 
$\Delta \ltap 0.015$ eV~
\footnote{Let us note that unless there is additional information
on the type of neutrino mass hierarchy or 
on the range of allowed values of $m_1$,
it will be in principle impossible 
to distinguish 
the case of IH spectrum from the 
one with partial normal hierarchy \cite{BPP1} 
($\deltasol \equiv \deltadue$ and $0.01 \ \eV < m_1 < 0.2 \ \eV$)
using only a measurement of \meff, as the predicted 
allowed ranges of \meff in the two cases overlap
(see Figs. 1 and 2).}. 
  
  The possibility of establishing  CP-
violation in the lepton sector due to Majorana CP-violating phases
has been studied in detail in ref.~\cite{PPR2}.
It was found that it requires quite accurate measurements
of \meff and of $m_1$, 
and holds only for a limited range of 
values of the relevant parameters.
For the IH and the QD spectra, which are 
of interest for this analysis,
the ``just CP-violation'' region~\cite{BPP1}
- an experimental point in this region
would signal unambiguously CP-violation
associated with Majorana neutrinos,
is larger for smaller values of 
$\cos 2 \theta_\odot$. 
As the present best fit values of 
\deltasol, \deltaatm and especially  
of $\sin^2 \theta_\odot$
are very similar to those
used in \cite{PPR2},
the conclusions reached in ref.
~\cite{PPR2} still hold.
More specifically, 
proving that CP-violation associated with
Majorana neutrinos takes place
requires, in particular, a relative 
experimental error on the measured value of 
\meff not bigger than (15 -- 20)\%,
a ``theoretical uncertainty'' in the value of
\meff due to an imprecise knowledge of the 
corresponding nuclear matrix elements
smaller than a factor of 2, a value of 
$\tan^2\theta_{\odot} \gtap 0.55$,
and values of the relevant Majorana
CP-violating phases ($\alpha_{21}$, 
$\alpha_{32}$) typically 
within the ranges of $\sim (\pi/2 - 3\pi/4)$ and
$\sim (5\pi/4 - 3\pi/2)$. 


\section{Conclusions}


  In the present Addendum, we have updated the predictions for
the effective Majorana mass in \betabeta-decay \meff
derived in ref.~\cite{PPSNO2bb} 
by taking into account the implications of the
recently announced results from the salt phase 
measurements of the SNO experiment.
The combined analyzes
of the solar neutrino data, including the latest
SNO results, lead to new relatively stringent
constraints on the solar neutrino mixing 
angle $\theta_{\odot}$:
i) the possibility of  $\cos2\theta_{\odot} = 0$
is excluded at more than 5 s.d.,
ii) the best fit value of 
$\cos2\theta_{\odot}$ is found to be  
$\cos2\theta_{\odot} = 0.40$, and
iii) at 95\% C.L. one has
$\cos2\theta_{\odot} \gtap 0.22$.
These new results firmly establish
the existence of significant
lower bounds on \meff 
in the cases of IH and QD neutrino mass spectra, 
which in turn has fundamental implications
for the searches for \betabeta-decay.
Using, e.g., the 90\% C.L. allowed
ranges of the values of the solar 
and atmospheric neutrino
oscillation parameters one finds
for the IH and QD spectra, respectively:
$\meff \gtap 0.010$ eV and 
$\meff \gtap 0.043$ eV.
The lower bounds obtained
utilizing the 95\% C.L.
allowed values of the parameters
do not differ substantially
from those given above.

  We have also updated the earlier results
in \cite{PPRSNO2bb,PPR2} on the possibilities
i) to discriminate
between the different types of 
neutrino mass spectrum
(NH vs IH, NH vs QD and IH vs QD), 
and ii) to get information 
about CP-violation
induced by the two Majorana CP-violating phases
in the PMNS mixing matrix, 
if a value $\meff \neq 0$ is measured in
the \betabeta-decay experiments
of the next generation.
The remarkable physics potential
of the future \betabeta-decay experiments
for providing quantitative information, 
in particular,
on the type of the neutrino mass spectrum and
on the CP-violation associated with 
Majorana neutrinos, can
be fully exploited
only if the values of 
the relevant $\betabeta-$decay nuclear 
matrix elements are known with 
a sufficiently small uncertainty.

\vspace{-0.4cm}
\section{Acknowledgments}
\vspace{-0.3cm}

 S.T.P. would like to thank D. Vignaud and the members
of APC Institute at College de France, Paris,
where parts of the work on the present study were done,
for kind hospitality. This work was supported in part by 
the Italian MIUR  under the program 
``Fenomenologia delle Interazioni Fondamentali'' 
(S.T.P.) and by the DOE Grant DE-FG03-91ER40662 and 
the NASA Grant ATP02-0000-0151 (S.P.).

\begin{figure}
\begin{center}
\epsfig{file=meffbf-3nu-add2003.epsi, height=16cm, width=10cm
}
\end{center}
\vspace{-3mm}
\caption{
The dependence of \meff on $m_1$ 
in the case of the LMA-I solution, 
for   $\deltasol = \Delta m_{21}^2$ and
$\deltasol = \Delta m_{32}^2$, and 
for the best fit values of the solar  neutrino 
oscillation parameters found in 
ref. \protect\cite{SNO3BCGPR}
and of \deltaatm in ref.~\protect\cite{Fogliatm0308055},
and fixed value of $\sin^2 \theta=0.0~(0.02)~[0.04]$ 
in the upper (middle) [lower] panel.
In the case of CP-conservation,
the allowed values of \meff are constrained 
to lie: for i) $\deltasol = \Delta m_{21}^2$ 
and the middle and lower panels (upper panel) -
{\it a)} on the lower thick solid lines
(on the lower thick solid line) if
$\eta_{21} = \eta_{31} = 1$,
{\it b)} 
on the  long-dashed lines (on the lower thick solid line) if
$\eta_{21} = - \eta_{31} = 1$,
{\it c)} 
on the dash-dotted lines (on the dash-dotted lines)
if $\eta_{21} = - \eta_{31} = - 1$,
{\it d)}
on the short-dashed lines (on the dash-dotted lines)
if $\eta_{21} = \eta_{31} = - 1$;
and for ii) $\deltasol = \Delta m_{32}^2$
(both panels) -
{\it a)}
on the upper thick solid line if
$\eta_{21} = \eta_{31} = \pm 1$,
{\it b)}
on the dotted lines
if $\eta_{21} = - \eta_{31} = \pm 1$.
In the case of CP-violation, the allowed regions
for \meff cover all the gray regions. 
Values of \meff in the dark gray regions 
signal CP-violation.} 
\label{fig:1}
\end{figure}

\begin{figure}
\begin{center}
\epsfig{file=meff90-3nu-add2003.epsi, height=16cm, width=10cm
}
\end{center}
\vspace{-3mm}
\caption{
The dependence of \meff on $m_1$ 
in the case of the LMA-I solution, 
for   $\deltasol = \Delta m_{21}^2$ and
$\deltasol = \Delta m_{32}^2$, and 
for the $90 \%$~C.L. allowed values 
of \deltasol and $\sin^2 \theta_\odot$ found in 
ref. \protect\cite{SNO3BCGPR}
and of \deltaatm in ref.~\protect\cite{Fogliatm0308055},
and a fixed value of $\sin^2 \theta = 0.0 (0.02) [0.04]$
in the upper (middle) [lower] panel.
In the case of CP-conservation,
the allowed values of \meff are constrained 
to lie: for i) $\deltasol = \Delta m_{21}^2$ 
and the middle and lower panels (upper panel) -
in the medium-gray and light-gray regions
{\it a)} between the two lower  thick solid lines 
(between the two lower  thick solid lines) if
$\eta_{21} = \eta_{31} = 1$,
{\it b)} 
between the two  long-dashed lines
(between the two lower  thick solid lines) if
$\eta_{21} = - \eta_{31} = 1$,
{\it c)} 
between the three thick  dash-dotted lines and the axes
(between the dash-dotted lines and the axes)
if $\eta_{21} = - \eta_{31} = - 1$,
{\it d)}
between the three thick  short-dashed lines and the axes
(between the dash-dotted lines and the axes)
if $\eta_{21} = \eta_{31} = - 1$;
and for ii) $\deltasol = \Delta m_{32}^2$
and the middle  and lower panels  (upper) -
in the  light-gray regions
{\it a)}
between the two upper  thick solid lines 
(between the two upper  thick solid lines) if
$\eta_{21} = \eta_{31} = \pm 1$,
{\it b)}
between the dotted  and 
the thin dash-dotted lines
(between the dotted  and 
the thick short-dashed lines)
if $\eta_{21} = - \eta_{31} =  1$,
{\it c)}
between the dotted  and 
the upper thick short-dashed lines
(between the dotted  and 
the thick short-dashed lines)
if $\eta_{21} = - \eta_{31} = - 1$.
In the case of CP-violation, the allowed regions
for \meff cover all the gray regions. 
Values of \meff in the dark gray regions 
signal CP-violation.} 
\label{fig:2}
\end{figure}


\begin{thebibliography}{99}
\baselineskip 10pt

 
\bibitem{PPSNO2bb} S. Pascoli and S.T. Petcov,
{\em Phys. Lett.} {\bf B544} (2002) 239.

\bibitem{SNO3} SNO Coll., S.N. Ahmed et al., arXiv:nucl-ex/0309004.


\bibitem{BiPet87} S.M.\ Bilenky and S.T.\ Petcov,
                {\em Rev.\ Mod.\ Phys.} \ {\bf 59} (1987) 67. 

\bibitem{ElliotVogel02} S.R. Elliot and P. Vogel, 
{\em Annu. Rev. Nucl. Part. Sci.} {\bf 52} (2002).

\bibitem{BPP1} S.M. Bilenky, S. Pascoli and S.T. Petcov,
               {\em Phys. Rev.} {\bf D64} (2001) 053010. 


\bibitem{SNO2} SNO Coll.,
               Q.R. Ahmad \textit{et al.}, 
{\em Phys. Rev. Lett.} {\bf 89} (2002) 011302 and 011301. 


\bibitem{Pont67} B. Pontecorvo, 
Zh. Eksp. Teor. Fiz. {\bf 53} (1967) 1717.


\bibitem{Cl98}   B.T. Cleveland {\em et al.}, 
                {\em  Astrophys. J.} {\bf 496} (1998) 505;
                Y.\ Fukuda {\em et al.},
               {\em  Phys.\ Rev.\ Lett.\ } {\bf 77} (1996) 1683;
                V.\ Gavrin, {\em  Nucl. Phys. Proc. Suppl.} {\bf 91} (2001) 36;
                W.\ Hampel {\em et al.},
               {\em  Phys.\ Lett.\ } {\bf B447} (1999) 127;
                M.\ Altmann {\em et al.},
               {\em  Phys.\ Lett.\ } {\bf B490} (2000) 16.


\bibitem{SKsol} Super-Kamiokande Coll.,
                Y. Fukuda {\em et al.}, 
              {\em  Phys.\ Rev.\ Lett. } {\bf 86} (2001) 5651. 



\bibitem{Pont46}
B. Pontecorvo, Chalk River Lab. report PD--205, 1946.

\bibitem{Davis68}
R. Davis, D.S. Harmer and K.C. Hoffman,
Phys. Rev. Lett. {\bf 20}, 1205 (1968); 
R. Davis, Proc. of the ``Neutrino~`72''~Int. Conference, Balatonfured,
    Hungary, June 1972 (eds. A. Frenkel and G.  Marx,  OMKDK-TECHNOINFORM,
    Budapest, 1972), p.5.



\bibitem{KamLAND} KamLAND Coll., K. Eguchi {\em et al.}, \
{\em Phys.\ Rev.\ Lett.}  {\bf 90} (2003) 021802.


\bibitem{fogli}
G.L. Fogli \textit{et al.}, 
{\em Phys.\ Rev.}  {\bf D67} (2003) 073002.
%
%
%
\bibitem{band}
A. Bandyopadhyay {\it et al.},
{\em Phys.\ Lett.}  {\bf B559} (2003) 121;
J.N. Bahcall, M.C. Gonzalez-Garcia, 
and C. Pe{\~n}a-Garay, hep-ph/0212147.

\bibitem{MarisSP02}
M.~Maris and S.~T.~Petcov,
Phys.\ Lett.\ B {\bf 534}, 17 (2002).


\bibitem{SNO3BCGPR} A. Bandyopadhyay {\it et al.}, hep-ph/0309174.

\bibitem{MarisSP00DN}
M.~Maris and S.~T.~Petcov,
Phys.\ Rev.\ D {\bf 62} (2000) 093006.


\bibitem{PPW} S. Pascoli, S.T. Petcov and L. Wolfenstein,
            {\em Phys. Lett.} {\bf B524} (2002) 319;
            S. Pascoli and  S.T. Petcov, hep-ph/0111203.
%

\bibitem{BGKP96} S.M. Bilenky et al., 
{\em Phys. Rev.} {\bf D56} (1996) .

\bibitem{Valle2003} M. Maltoni {\it et al.}, hep-ph/0309130. 
 
\bibitem{bala2003} A.B. Balantekin and H. Y\:ksel,
hep-ph/0309079; 
G.L. Fogli {\it et al.}, hep-ph/0309100;

\bibitem{SNO3Milano} P. Aliani et al., hep-ph/0309156.


\bibitem{CHOOZPV} M.\ Apollonio \textit{et al.}, 
                 {\em Phys. Lett. }{\bf B466} (1999) 415;
F. Boehm \textit{et al.}, 
{\em Phys. Rev.} {\bf D62} (2000) 072002.


\bibitem{SKatmo03} Super-Kamiokande Coll., 
Y. Hayato \textit{et al.}, Talk given at the Int. EPS Conference
on High Energy Physics, July 17 - 23, 2003, Aachen, Germany.


\bibitem{FogliatmKamL} G.L. Fogli \textit{et al.}, 
{\em Phys. Rev.} {\bf D67} (2003) 093006.


\bibitem{K2K} K2K Coll., M.H. Ahn et al., 
{\em Phys.\ Rev.\ Lett.\ }  {\bf 90} (2003)  041801.


\bibitem{Fogliatm0308055} G.L. Fogli et al., hep-ph/0308055.


\bibitem{BNPChooz} S.M. Bilenky, D. Nicolo and S.T. Petcov, 
{\em Phys. Lett.} {\bf B538} (2002) 77.

\bibitem{BPont57} B. Pontecorvo, 
                  {\em Zh. Eksp. Teor. Fiz.} {\bf 33}, 549 (1957),
                and {\bf 34}, 247 (1958);
Z. Maki, M. Nakagawa and S. Sakata, 
{\em Prog. Theor. Phys.} {\bf 28} (1962) 870.


\bibitem{BHP80} S.M.\ Bilenky \textit{et al.},
              {\em  Phys.\ Lett.}  {\bf B94} (1980) 495.

\bibitem{Doi81} M.~Doi \textit{et al.},
{\em Phys. Lett.}  \textbf{B102} (1981) 323.

\bibitem{SPAS94} S.T. Petcov and A.Yu. Smirnov,
                   {\em Phys. Lett.}  \textbf{B322} (1994) 109;
S.M. Bilenky \textit{et al.}, {\em Phys.\ Lett.} 
{\bf B465} (1999) 193.

\bibitem{PPRSNO2bb} S. Pascoli, S.T. Petcov and
W. Rodejohann,
{\em Phys. Lett.} {\bf B558} (2003) 141.


\bibitem{PPR2}  S. Pascoli, S.T. Petcov and
W. Rodejohann, {\em Phys. Lett.} {\bf B549} (2002) 177. 


\bibitem{PPVenice03} S. Pascoli and S.T. Petcov,
hep-ph/0308034.


\bibitem{Carlosbb03} H. Murayama and C. Pe{\~n}a-Garay, hep-ph/0309114. 



\end{thebibliography}
\end{document}